\documentclass[aps,pra,showpacs,groupedaddress,superscriptaddress,twocolumn,floatfix]{revtex4}
\usepackage{amsmath}
\usepackage{amssymb}
\usepackage[dvips]{graphics,color}
\usepackage{graphicx}
\usepackage{latexsym}

\newcommand{\be}{\begin{equation}}
\newcommand{\ee}{\end{equation}}
\newcommand{\beqn}{\begin{eqnarray}}
\newcommand{\eeqn}{\end{eqnarray}}
\newcommand{\beqnn}{\begin{eqnarray*}}
\newcommand{\eeqnn}{\end{eqnarray*}}

\input{tcilatex}
\begin{document}

\title{How to check the one-count operator experimentally}
\author{A. V. Dodonov}
\email{adodonov@df.ufscar.br}
\author{S. S. Mizrahi}
\email{salomon@df.ufscar.br} \affiliation{Departamento de
F\'{\i}sica, CCET, Universidade Federal de S\~{a}o Carlos, Via
Washington Luiz km 235, 13565-905, S\~ao Carlos, S\~ao Paulo,
Brazil}
\author{V. V. Dodonov}
\email{vdodonov@fis.unb.br} \affiliation{Instituto de F\'{\i}sica,
Universidade de Bras\'{\i}lia, PO Box 04455, 70910-900,
Bras\'{\i}lia, Distrito Federal, Brazil}
\date{\today }

\begin{abstract}
We propose an experimental scheme to probe the form of one-count
operation used in the theory of continuous photodetection in
cavities. Two main steps are: 1) an absorption of a single photon by
an atom passing through a high-Q cavity containing electromagnetic
field in a thermal or coherent state, 2) a subsequent measurement of
the photon statistics in the new field state arising after the
photon absorption. Then comparing the probabilities of finding 0 and
1 photons in the initial and final states of the field, one can make
conclusions on the form of the one-count operation. This method can
be readily applied in the microwave cavity QED with present
technology.
\end{abstract}

\pacs{03.65.Ta, 42.50.Ar, 42.50.Lc} \maketitle

It is well known \cite{Gla130} that the probability of absorbing one
photon per unit time from a quantized electromagnetic field is
proportional to the average value of the ordered product of the
negative and positive frequency electric field operators in the
given quantum state of the field. In the simplest case of the
single-mode field, this probability can be written in
terms of the standard bosonic lowering and raising operators $\hat{a}$ and $%
\hat{a}^{\dagger }$, satisfying the commutation relation $[\hat{a},\hat{a}%
^{\dagger }]=1$, as 
\begin{equation}
p_{ab}=\gamma \mathrm{Tr}\left( \hat{a}\hat{\rho}_{i}\hat{a}^{\dagger
}\right)
  \label{p-ab}
\end{equation}%
where $\hat{\rho}_{i}$ is the statistical operator of the field \emph{before
absorption\/} and $\gamma $ is a coefficient with the dimensionality
s$^{-1}$. Due to an interaction with a \textquoteleft
detector\textquoteright\ (which absorbs a photon), the field
makes a \textquoteleft quantum jump\textquoteright\ to a new
state, which can be described mathematically by an action of the
\emph{one-count operator} (OCO) $\hat{J}$ as  \cite{SD}
\begin{equation}
\hat{\rho}_{f}=\hat{J}\hat{\rho}_{i}/\mathrm{Tr}(\hat{J}\hat{\rho}_{i})
\label{def-J}
\end{equation}%
where $\hat{\rho}_{f}$ is the statistical operator of the field immediately
after the absorption of one photon.
Operator $\hat{J}$ is frequently called also
 \emph{quantum jump superoperator\/} (QJS).
 However, this term is usually associated with random processes
 and the so called `quantum trajectories approach'
 (see e.g. \cite{carm,Plen}).
 In order to avoid
 a confusion, we shall use the term OCO throughout the paper.

The hermiticity of operator $\hat{\rho}_{f}$ can be ensured
if one uses the decomposition
\begin{equation}
\hat{J}\hat{\rho}\equiv \hat{O}\hat{\rho}\hat{O}^{\dagger }
\end{equation}%
where $\hat{O}$ is some \textquoteleft lowering\textquoteright\
operator responsible for the subtraction of one photon from the field.
Obviously, the explicit form of operators $\hat{J}$ or $\hat{O}$ depends on
the details of the interaction between the field and a detector, and
concrete calculations based on different models were performed by many
authors since the 1960s \cite{KK64,Moll68,Imo90,Ag94} (other references can
be found in \cite{PL00}).
A very common form of OCO, first proposed in \cite{SD} and considered for
applications in quantum-counting quantum nondemolition (QND) measurements in
\cite{MilWal84}, consists in the identification $\hat{O}=\hat{a}$
(we shall refer to it as \textquoteleft A-model\textquoteright):
\begin{equation}
\hat{J}^{A}\hat{\rho} _{i}=\hat{a}\hat{\rho}_{i}\hat{a}^{\dagger }
\label{J-A}
\end{equation}%
Such a form seems quite natural, if not obvious,
in view of equation (\ref{p-ab}). However,
we would like to emphasize that this choice is, as a matter of fact, \emph{%
intuitive} (\emph{phenomenological\/}), although it can be derived from some
\textquoteleft microscopical\textquoteright\ models under certain
assumptions \cite{Imo90,DMD05}, where the most important are the weak
coupling and short interaction time limits. Nonetheless, if these
assumptions are replaced by others, one can obtain different operators $\hat{%
J}$. In particular, the OCO $\hat{J}^{n}\hat{\rho}_{i}=\hat{n}\hat{\rho}_{i}%
\hat{n}$, where $\hat{n}\equiv \hat{a}^{\dagger }\hat{a}$ is the photon
number operator, was considered in \cite{Ueda92} in connection with
continuous quantum nondemolition measurements of photon number. A family of
OCO based on the \textquoteleft nonlinear lowering
operators\textquoteright\ of the form $\hat{O}=(1+\hat{n})^{-\beta }\hat{a%
}$ was derived in Ref. \cite{DMD05}. Its special case with $\beta
=1/2$ corresponds to the so-called \textquoteleft E-model\textquoteright,
which was proposed within the frameworks
of phenomenological considerations in \cite{benaryeh,OMD-JOB}:
\begin{equation}
\hat{J}^{E}\hat{\rho}_{i}=\hat{E}_{-}\hat{\rho}_{i}\hat{E}_{+},\quad \hat{E}%
_{-}\equiv (1+\hat{n})^{-1/2}\hat{a}.  \label{J-E}
\end{equation}%
The operator $\hat{E}_{-}$ is known under the name \textquoteleft
exponential phase operator\textquoteright\ \cite{Suss,CarNi,Wun,Kast}.

In some special cases, e.g., if a detector is
a resonant two-level atom passing through a cavity,
one can {\em deduce\/} an exact form of the
one-count operator, using some known atom--field interaction
Hamiltonian. Indeed, if one can describe
the interaction by means of the Jaynes--Cummings model, then
the \emph{exact} form of the OCO is \cite{job05}
\begin{equation}
\hat{J}^H\rho_i=\sin(y\sqrt{\hat{n}+1})\hat{E}_-\rho_i
\hat{E}_+\sin(y\sqrt{\hat{n}+1})
\label{o}
\end{equation}
where $y=gt$, $t$ being the atom transit time through the cavity and
$g$ the atom-field coupling constant related to the Rabi frequency.
We shall call operator (\ref{o}) the `H-model'.
Common values in cavity QED in the microwave regime are
(see, e.g., table I in \cite{blais}):
 $g\sim 100$kHz, $t\sim 100\mu$s, so $y\sim 1-10$.
Notice that the exact OCO (\ref{o}) is a bounded superoperator, as
expected from physical point of view.

Although the OCO in the form (\ref{J-A}) was used \emph{ad hoc} for more
than three decades in numerous papers devoted to different applications \cite%
{PL00}, it seems that its validity was never verified in direct experiments.
However, such a verification cannot be considered as unnecessary for several
reasons. First, it is possible that in some realistic situations, the
approximations under which the \emph{phenomenological\/} operator (\ref{J-A}%
) was derived can fail. Second, since $\hat{J}^{A}$ is an unbounded
operator, some inconsistencies in the theoretical treatment appear (they
were noticed already in the original paper \cite{SD}; see also \cite%
{OMD-JOB,job05,DMD06}). Third, applying (\ref{J-A}) to some states,
one arrives at predictions which look counterintuitive, thus
deserving an experimental verification.

For example, it is easy to check that if the mean number of photons in the
state $\hat{\rho}_{i}$ (before the detection of one photon) was $\langle
\hat{n}\rangle _{i}$, then the mean number of photons in the state $\hat{\rho%
}_{f}$ (\ref{def-J}) with operator (\ref{J-A}) must be \cite%
{Ueda90,Lee93,MD02}
\begin{equation}
\langle \hat{n}\rangle _{f}=\langle \hat{n}^{2}\rangle _{i}/\langle
\hat{n}\rangle _{i}-1\equiv \langle \hat{n}\rangle _{i}+Q
\label{new-n}
\end{equation}%
where $Q$ is the known Mandel's $Q$-factor describing the type of photon
statistics in the initial state $\hat{\rho}_{i}$. Only for the initial Fock
states one has $\langle \hat{n}\rangle _{f}=\langle \hat{n}\rangle _{i}-1$,
whereas equation (\ref{new-n}) yields $\langle \hat{n}\rangle _{f}=2\langle \hat{n%
}\rangle _{i}$ for the initial thermal state and $\langle \hat{n}\rangle
_{f}>2\langle \hat{n}\rangle _{i}$ for the initial squeezed vacuum state. In
contrast, using OCO in the form (\ref{J-E}) one obtains instead of (\ref%
{new-n}) the formula
\begin{equation}
\langle \hat{n}\rangle _{f}=\frac{\langle \hat{n}\rangle _{i}}{1-\chi _{0}}%
-1, \qquad \chi _{0}\equiv \langle 0|\hat{\rho}_{i}|0\rangle
\label{new-nE}
\end{equation}%
where $\chi _{0}$ is the probability of occupation of the vacuum state in
the initial state $\hat{\rho}_{i}$. In particular, for the thermal state
equation (\ref{new-nE}) yields
$\langle \hat{n}\rangle _{f}=\langle \hat{n}\rangle_{i}$.

The aim of this article is to show how the form of the OCO can be verified
by detecting single photons in high-Q cavities (where one can use the
single-mode approximation for the quantized electromagnetic field). We are
inspired by the recent progress in experiments described in
 \cite{exp06}. The scheme that we propose employs both destructive and nondemolition
measurements, that can be realized with the present available technology
\cite{exp06,Rai01}.

In quantum nondemolition experiments realized recently (based on a proposal
made in \cite{haro}), the Rydberg atoms, initially prepared in the ground
state $|g\rangle $ of an effective two-level configuration, were sent through
an interferometer composed of a high-Q cavity (with the damping time $\sim
0.1\,$s) and resonant classical fields. On the exit they were detected by a
state selective field ionization detector. Besides, the experiments were
performed under the conditions where the mean number of photons in the
cavity was much smaller than unity. In such a case, due to the nondemolition
nature of measurements (because the cavity field eigenfrequency is chosen in
such a way that the atomic transitions are \emph{out of resonance\/} with
the field), if the atom is detected in the excited state $|e\rangle $, then
one may conclude that there is only one photon in the cavity, so the field
state within the cavity is projected into the 1-photon state. Similarly, if
the atom is detected in the state $|g\rangle $, this means that there are no
photons in the cavity, and the field state is the vacuum state. If one sends
more atoms through the cavity, the outcomes of the measurements will be the
same and the state within the cavity will not be altered. In rare cases when
there is more than 1 photon in the cavity, the atom will be in a
superposition of states $|g\rangle $ and $|e\rangle $ after passing through
the cavity, so in consecutive measurements the outcome will not be always
the same, but will alternate probabilistically between $|g\rangle $ and $%
|e\rangle $. Thus, using consecutive nondemolition measurements, an
experimenter can distinguish between 0, 1 and more than 1 photon in the
cavity.

Our experimental proposal is based on the assumption that one can
prepare a field state $\hat{\rho}_{i}$ in the cavity with known
statistical properties. Actually, we have in mind either a thermal
or a coherent state with a small mean photon number $\langle
n\rangle _{i}<5$, in order to ensure a negligibly small influence of
multiphoton Fock states. The methods of preparation of such
\textquoteleft classical\textquoteright\ states seem to be
well known. (Note that the Fock states themselves cannot distinguish
between the OCO's -- one needs superpositions or mixtures of these
states.) If the nature of the state is known, then it can be
characterized by measuring the ensemble probabilities $\chi _{0}$
and $\chi _{1}$ of having initially 0 and 1 photons. So, the first
step of the experiment is the QND measurement of the photon
statistics in the initial state. After this, one should send through
the cavity an atom in the ground state of another effective
two-level configuration, tuned \emph{in resonance\/} with the cavity
mode (e.g., using Rydberg atoms, whose quantum states are different
from those used in the first step), in order to change the quantum
state of the field due to the absorption of one photon. If the atom
absorbs a photon (which is signaled by a detection of atom in the
excited state), this means that the field state makes a quantum jump
to the state $\hat{\rho}_{f}$, whose statistical properties are
determined by the form of OCO $\hat{J}$. Consequently, measuring the
probabilities $P_{n}=\langle n|\hat{\rho}_{f}|n\rangle $ of finding
$n$ photons in the state $\hat{\rho}_{f}$ after the quantum jump and
comparing the results with theoretical predictions, one can verify
the form of $\hat{J}$. It is sufficient to measure only the
probabilities $P_{0}$ and $P_{1}$.

The predictions for the A-model are as follows,
\begin{equation}
P_{n}^{A}=\frac{\langle n|\hat{a}\hat{\rho}_{i}\hat{a}^{\dagger }|n\rangle }{%
\mathrm{Tr}\left[ \hat{a}^{\dagger }\hat{a}\hat{\rho}_{i}\right] }=\frac{%
(n+1)!}{\langle \hat{n}\rangle _{i}}\chi _{n+1},
\end{equation}%
where $\chi _{n}=\langle n|\rho _{i}|n\rangle $. Analogously, for
the E-model we have%
\begin{equation}
P_{n}^{E}=\frac{\chi _{n+1}}{1-\chi _{0}}
\end{equation}
and for the H-model
\begin{equation}
P_{n}^{H}=\frac{\sin^2(y\sqrt{n+1})\chi
_{n+1}}{\langle\sin^2(y\sqrt{n})\rangle_i}.
\end{equation}

\begin{figure}[t]
\begin{center}
\includegraphics[width=0.46\textwidth]{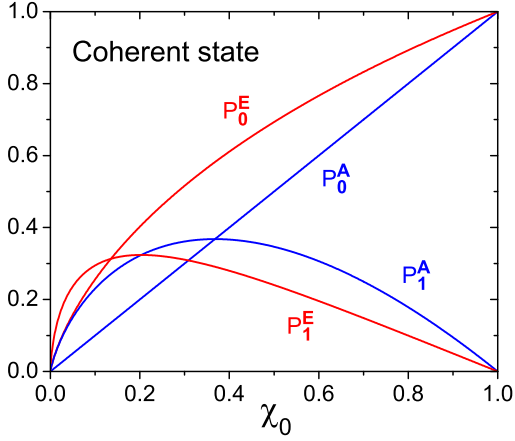}
{\caption{Probabilities of finding 0 and 1 photons after the quantum
jump from the initial coherent state, characterized by the initial
probability of having zero photons $\chi_0$. The superscripts A and
E correspond to predictions of A-model and E-model, respectively.}
\label{fig-coh}}
\end{center}
\end{figure}
\begin{figure}[t]
\begin{center}
\includegraphics[width=0.46\textwidth]{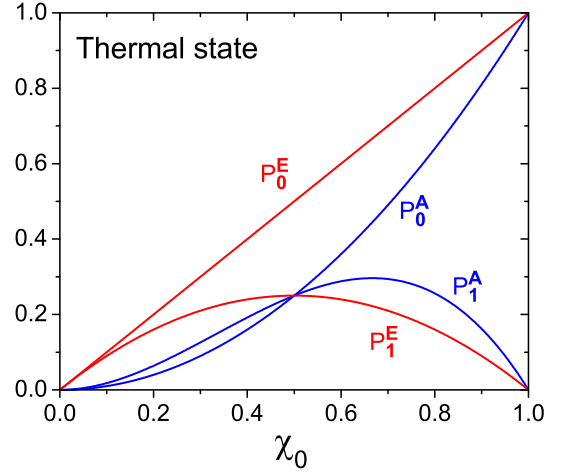}
{\caption{The same as in figure \ref{fig-coh}, but for the initial
thermal state.} \label{fig-therm}}
\end{center}
\end{figure}

Thus, we see that the resulting probabilities are fundamentally
different. Let us illustrate these different behaviors for the A- and
E-models for two different initial states
(for H-model the expressions are more lengthy, so we do not
put them here).

\textbf{(a)} For the thermal state (which is an eigenstate of superoperator $%
\hat{J}^{E}$) with the mean photon number $\bar{n}$ we have%
\begin{equation}
\chi _{n}=\frac{\bar{n}^{n}}{(\bar{n}+1)^{n+1}}=\chi _{0}\left( 1-\chi
_{0}\right) ^{n}
\label{chi-nther}
\end{equation}
so we obtain $P_n^E=\chi_n$,%
\begin{equation*}
P_{0}^{A}=\chi _{0}^{2}, \qquad P_{1}^{A}=2\chi _{0}^{2}\left( 1-\chi
_{0}\right) .
\end{equation*}%

\textbf{(b)} For the coherent state (an eigenstate of $\hat{J}^{A}$)
with
\begin{equation*}
\chi _{n}=e^{-\bar{n}}\frac{\bar{n}^{n}}{n!}=\chi _{0}\frac{(-\ln
\,\chi _{0})^{n}}{n!}
\end{equation*}%
we have $P_n^A=\chi_n$,%
\begin{equation*}
P_{0}^{E}=\frac{\chi _{0}(-\ln \,\chi _{0})}{1-\chi _{0}}, \qquad
P_{1}^{E}=%
\frac{\chi _{0}(-\ln \,\chi _{0})^{2}}{2(1-\chi _{0})}.
\end{equation*}%
We see that $P_{1}^{E}$ is twice smaller than $P_{1}^{A}$ if
$1-\chi_0\ll 1$,
for both initial coherent and thermal quantum states.

\begin{figure}[ht]
\begin{center}
\includegraphics[width=0.46\textwidth]{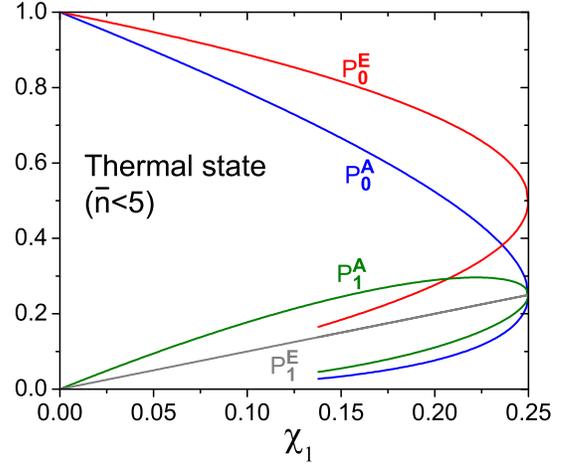}
{\caption{The same as in figure \ref{fig-therm}, but as functions of
the probability $\chi_1$.} \label{fig-therm-chi1}}
\end{center}
\end{figure}
\begin{figure}[t]
\begin{center}
\includegraphics[width=0.46\textwidth]{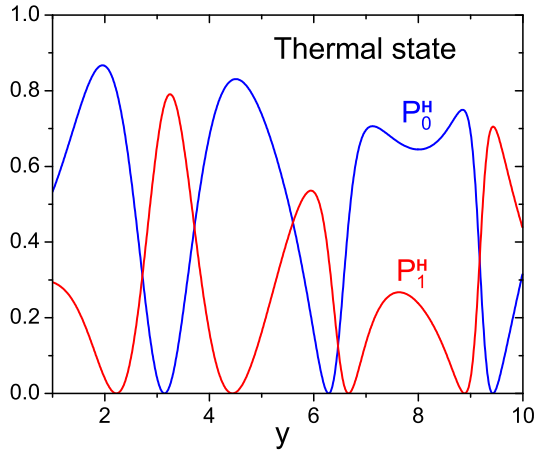}
{\caption{$P_n^H$ for the H-model as functions of $y$ for
$\chi_0=0.6$.} \label{fig4}}
\end{center}
\end{figure}

In figures \ref{fig-coh} and \ref{fig-therm} we plot $P_{0}$ and $P_{1}$ as
function of $\chi _{0}$ for A and E models and the both states. In figure \ref%
{fig-therm-chi1} we plot the same probabilities as functions of $\chi _{1}$
for the initial thermal state (in the case of $\bar{n}<5$). We choose $%
\chi_{0}$ and $\chi_{1}$ as possible independent variables,
because these quantities
can be determined experimentally in the most direct way.
Two branches in figure \ref{fig-therm-chi1} are the consequence
of two signs in the dependence $\chi_0(\chi_1)$: solving equation
(\ref{chi-nther}) with respect to $\chi_0$ for $n=1$ one obtains
\[
\chi_0= 1/2 \pm \sqrt{1/4 - \chi_1}.
\]
The upper sign should be chosen if $\bar{n} <1$ and the lower sign
corresponds to $\bar{n} >1$.

In the case of initial thermal states, the values of $\chi_{0}$ and
$\chi_{1}$ can be
varied by changing the temperature of the cavity or by some other means \cite%
{Nog99}. Before passing any atom, the mean number of (initial) thermal
photons in the set-up described in  \cite{exp06,Rai01,haro,Nog99}
varied from $0.7$ to $0.1$. This range of temperatures corresponds to the
variations of $\chi _{0}$ from $0.6$ to $0.9$ and $\chi _{1}$ from $0.24$ to
$0.09$. Figures \ref{fig-therm} and \ref{fig-therm-chi1} show that these are
just the intervals where the functions $P_{k}^{A}(\chi _{j})$ and $%
P_{k}^{E}(\chi _{j})$ are quite distinguishable from each other
($k,j=0,1$). Moreover, for $\chi_{1}=0.1$, the probability of
detecting more than one photon becomes less than $0.01$, and the
scheme described in \cite{exp06} is quite reliable.

In the H-model the OCO depends on the parameter $y$, i.e., the atom
transit time. Thus, the resulting probabilities $P_n^H$ oscillate as
functions of transit time, attaining zero values for certain values
of $y$. In figure \ref{fig4} we plot functions $P_0^H(y)$ and
$P_1^H(y)$ for the thermal state with $\chi_0=0.6$ and $y$ ranging
from 1 to 10, corresponding to achievable values in microwave cavity
QED experiments. Such a peculiar behavior of probabilities as
functions of the transit time could also be checked experimentally.
Consequently, by performing ensemble experiments in an accessible
interval of temperatures one can easily verify which one of the
OCO's holds, or whether neither of them is observed in practice.

Concluding, we are proposing a simple scheme of an experiment, which
could decide in an unambiguous way the form of the one-count
operator. This scheme only needs a cavity with initial thermal or
coherent state of the electromagnetic field containing a small mean
number of photons. The available experimental level seems to be
quite sufficient for this purpose. This method can also be applied
to other physical systems in which one can perform both destructive
and QND (or instantaneous destructive photon number) measurements.

AVD and SSM acknowledge financial support from FAPESP (SP, Brazil,
Contract No. 04/13705-3). SSM and VVD acknowledge partial financial
support from CNPq (DF, Brazil).


\begin{thebibliography}{99}
\bibitem{Gla130} Glauber R J 1963 {\em Phys. Rev.} \textbf{130} 2529

\bibitem{SD} Srinivas M D and  Davies E B 1981 {\em Opt. Acta\/}
{\bf28} 981

\bibitem{carm} Carmichael H J 1997 {\em  Phys. Rev.} A \textbf{56}
5065

\bibitem{Plen}  Plenio M B and  Knight P L 1998
{\em Rev. Mod. Phys.} \textbf{70} 101


\bibitem{KK64} Kelley P L and  Kleiner W H 1964 {\em Phys. Rev.} {\bf136} A316

\bibitem{Moll68}  Mollow B R 1968 {\em Phys. Rev.} {\bf168} 1896

\bibitem{Imo90} Imoto N, Ueda M and Ogawa T 1990
{\em Phys. Rev.} A {\bf 41} 4127


\bibitem{Ag94} Agarwal G S, Graf M, Orszag M, Scully M O and Walther H 1994
{\em Phys. Rev.} A {\bf 49} 4077


\bibitem{PL00} Pe\v{r}inov\'a V and Luk\v{s} A 2000
{\em Progress in Optics\/} {\bf 40} ed E Wolf (Amsterdam: Elsevier) p 115


\bibitem{MilWal84} Milburn G J and  Walls D F 1984
{\em Phys. Rev.} A {\bf 30} 56


\bibitem{DMD05} Dodonov A V, Mizrahi S S  and Dodonov V V 2005
{\em  Phys. Rev.} A \textbf{72} 023816

\bibitem{Ueda92} Ueda M, Imoto N, Nagaoka H and Ogawa T 1992
{\em Phys. Rev.} A {\bf 46} 2859

\bibitem{benaryeh} Ben-Aryeh Y and Brif C 1995 {\em Preprint\/}
quant-ph/9504009 (unpublished)


\bibitem{OMD-JOB} de Oliveira M C, Mizrahi S S and Dodonov V V 2003
{\em J. Opt. B: Quantum. Semiclass. Opt.} {\bf 5} S271

\bibitem{Suss} Susskind L and Glogower J  1964 {\em Physics\/}  {\bf 1} 49


\bibitem{CarNi} Carruthers P and Nieto M 1968
{\em Rev. Mod. Phys.}  {\bf 40} 411

\bibitem{Wun} W\"unsche A 2001
{\em J. Opt. B: Quantum. Semiclass. Opt.} {\bf 3} 206

\bibitem{Kast} Kastrup H A 2003,
{\em Fortschr. Phys.\/} {\bf 51} 975

\bibitem{job05} Dodonov A V, Mizrahi S S  and Dodonov V V 2005
{\em J. Opt. B: Quantum. Semiclass. Opt.} {\bf 7} 99

\bibitem{blais} Blais A,  Huang R-S,  Wallraff A, Girvin S M
and  Schoelkopf R J 2004 {\em Phys. Rev.} A \textbf{69} 062320


\bibitem{DMD06} Dodonov A V, Mizrahi S S  and Dodonov V V 2006
{\em  Phys. Rev.} A \textbf{74} 033823 \\
Dodonov A V, Mizrahi S S  and Dodonov V V 2007
{\em  Phys. Rev.} A \textbf{75} 013806


\bibitem{Ueda90} Ueda M, Imoto N and Ogawa T 1990
{\em Phys. Rev.} A {\bf 41} 3891


\bibitem{Lee93} Lee C T 1993 {\em Phys. Rev.} A  {\bf 48} 2285


\bibitem{MD02} Mizrahi S S  and Dodonov V V 2002
{\em J. Phys. A: Math. Gen.} \textbf{35} 8847

\bibitem{exp06} Gleyzes S, 
 Kuhr S,  Guerlin C,  Bernu J,  Del\'eglise S,  Hoff U B,  Brune M,
Raimond J-M and  Haroche S 2007 {\em Nature} \textbf{446} 297 {\em
Preprint\/} quant-ph/0612031

\bibitem{Rai01} Raimond J-M, Brune M and  Haroche S 2001
{\em Rev. Mod. Phys.} \textbf{73} 565

\bibitem{haro} Brune M, Haroche S, Lef\`evre V, Raimond J-M and
Zagury N 1990 {\em Phys. Rev. Lett.} \textbf{65} 976 \\
Brune M, Haroche S, Raimond J-M, Davidovich L and Zagury N 1992
{\em Phys. Rev.} A \textbf{45} 5193

\bibitem{Nog99}  Nogues G,  Rauschenbeutel A,  Osnaghi S,
Brune M, Raimond J-M and  Haroche S 1999
{\em Nature\/} \textbf{400} 239

\end{thebibliography}
\end{document}